# A Ransomware Classification Framework Based on File-Deletion and File-Encryption Attack Structures

Aaron Zimba
*Department of Computer Science and Information Technology*
*Mulungushi University*
*Kabwe*
*Zambia*
azimba@mu.ac.zm

Mumbi Chishimba
*Department of Information Technology*
*National Institute of Public Administration (NIPA)*
*Lusaka, Zambia*
chishimba.mumbi@gmail.com

Sipiwe Chihana
*Center for Information and Communications Technology*
*Northrise University*
*Ndola, Copperbelt*
*Zambia*
Sipiwechihana09@gmail.com

*Abstract*— Ransomware has emerged as an infamous malware that has not escaped a lot of myths and inaccuracies from media hype. Victims are not sure whether or not to pay a ransom demand without fully understanding the lurking consequences. In this paper, we present a ransomware classification framework based on file-deletion and file-encryption attack structures that provides a deeper comprehension of potential flaws and inadequacies exhibited in ransomware. We formulate a threat and attack model representative of a typical ransomware attack process from which we derive the ransomware categorization framework based on a proposed classification algorithm. The framework classifies the virulence of a ransomware attack to entail the overall effectiveness of potential ways of recovering the attacked data without paying the ransom demand as well as the technical prowess of the underlying attack structures. Results of the categorization, in increasing severity from CAT1 through to CAT5, show that many ransomwares exhibit flaws in their implementation of encryption and deletion attack structures which make data recovery possible without paying the ransom. The most severe categories CAT4 and CAT5 are better mitigated by exploiting encryption essentials while CAT3 can be effectively mitigated via reverse engineering. CAT1 and CAT2 are not common and are easily mitigated without any decryption essentials.

## I. INTRODUCTION

Since the invention of the Internet, cyber-crime has continued to grow [1] with attackers employing more innovative ways to attain proceeds of cyber-crime. Since the motivation behind most cyber-crime is monetary gain (excluding cyber espionage and hacktivism), the challenge mainly has been to seamless collect the associated monetary proceeds without a trace. The invention of Bitcoin seems to be a dream come true for cyber criminals due to the anonymity provided by the Bitcoin system [2]. As such, attackers eschewing data exfiltration attacks for less tedious attacks with a high turnover. One such attack is ransomware where the attacker takes hostage of the victim's data without the need to exfiltrate it at all. In a ransomware attack, the attacker uses robust and resilient encryption to make the target data inaccessible without the appropriate decryption keys [3]. Furthermore, the attacker demands a ransom in Bitcoins and usually the victim is left with a binary option of whether to pay or not to. This has seen some victims part away with over a million dollars in a single attack [4]. As such, the ransomware business model is a multi-billion lucrative industry in the cyber-crime landscape which is growing each day [5] with criminal business concepts such as Ransomware-as-a-service [6]. The popularity of ransomware is echoed by Interest Over Time (IOT) as shown in figure 1 below.

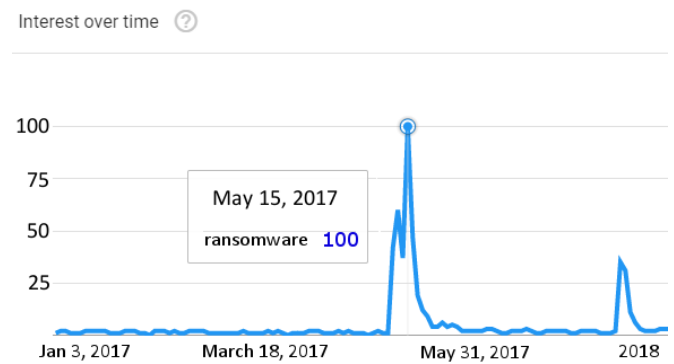

Figure 1. Ransomware attacks IOT. [7]

Sadly, the myths and inaccuracies around ransomware continue to deepen. This has caused victims to make uninformed decisions upon a ransomware attack. Depending on the underlying attack structures, some ransomware attacks can be mitigated and the data recovered without paying the ransom. Unfortunately, some victims have had to pay ransom demands when data could be recovered without honoring the ransom demand [8], as was with the major ransomware attack of 2017 depicted in figure 1. As such, knowledge of a ransomware's attack structure is vital to the mitigation thereof.

In light of the aforesaid, this paper evaluates attack methodologies of a ransomware attack: the underlying file-deletion and file-encryption attack structures. In the former, we uncover the data recovery-prevention techniques and in the latter, we uncover the associated cryptographic attack models. The deeper comprehension of potential flaws and inadequacies exhibited in these attack structures form the basis of the overall objective. This enables the provision of enough technical information before making a hasty decision to pay a ransom which might result into not only financial loss but loss of access to the attacked files if decryption is not possible by the attacker. We present a threat and attack model which is representative of a typical ransomware attack process from which we derive the ransomware categorization framework based on a proposed classification algorithm. The framework classifies the virulence of a ransomware attack to entail the overall effectiveness of potential ways of recovering the attacked data without paying the ransom demand as well as the technical prowess of the underlying attack structures.

The rest of the paper is organized as follows: Section II discusses the taxonomy and the threat model with the associated attack structures while Section III presents the proposed classification framework. The methodology and approach are presented in Section IV while classification results and the analysis thereof are brought forth in Section V. The conclusions of the paper are drawn Section VI.

## II. TAXONOMY, THREAT MODEL AND ATTACK STRUCTURES

### A. Ransomware Attacks taxonomy

There are several factors that affect the categorization of ransomware attacks. We categorize the attacks based on the following characteristics; target platform, cryptosystem used, severity of loss of data and attack structure. This categorization is not dependent on the underlying infection vectors. The result of the categorization is shown in figure 2 below.

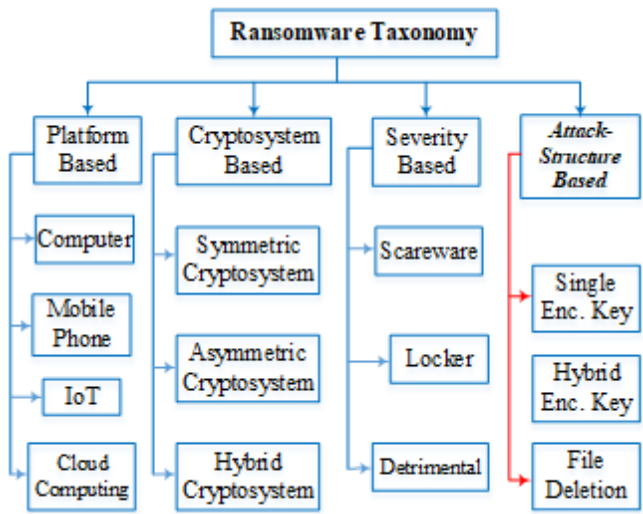

Figure 2. A taxonomy of ransomware attacks

Based on the target platform, ransomware can be made to target computers, mobile phones, Internet of Things devices or cloud computing. Computer based ransomware is the most prevalent owing to wide attack surface and ease of implementation. Android-based ransomware dominates the mobile phone landscape [9] whilst Windows-based IoT devices have been found to be the most susceptible. Cloud computing is an emerging niche for ransomware attacks [10] but the most effective ransomware attack in this domain have been targeted attacks [11] and not indiscriminate. Owing to the disparities in the target platform, we do not use it as a basis for formulating our proposed classification framework. Regarding the cryptosystems used in ransomware today, they can be classified as those that use symmetric, asymmetric or hybrid cryptosystem. In symmetric cryptosystem, the same key is used to encrypt and decrypt the target data. As such the attacker has the classic challenge of secure key management [12]. To overcome the challenge of key management, asymmetric crypto systems are used where the public key is used to encrypt the target data whereas the corresponding private key retained by the attacker is used for decryption. The challenge in asymmetric encryption is that it is slow. Hybrid encryption is adopted as an alternative as it utilizes the speed of symmetric cryptosystems and the resilience of asymmetric cryptosystems. We use some facets of encryption as one of the bases of our framework considering that encryption is at the core of the ransomware business model. In terms of attack structures, ransomware attacks can be categorized into those that use a single key (either a symmetric key or a public), hybrid key (use of both symmetric and public key) and file deletion which either overwrites the original file after encryption or primitively deletes it. We incorporate this characteristic in our framework as it is pivotal to data recovery after an attack. In terms of severity of the damaged caused, scareware type of ransomware doesn't damage or delete the files, it just obfuscates them in one form or another. Locker ransomware usually locks the system login or the boot menu. As such, offline mitigation is effective against this attack category. Detrimental ransomware is one that both encrypts the target files and deletes the remnant original files after encryption.

### B. Threat Model and Attack Structures

We now evaluate the threat model and attack structures based on the two selected categories from the taxonomy. The diagram below in figure 2 depicts the resultant threat model and attack structures.

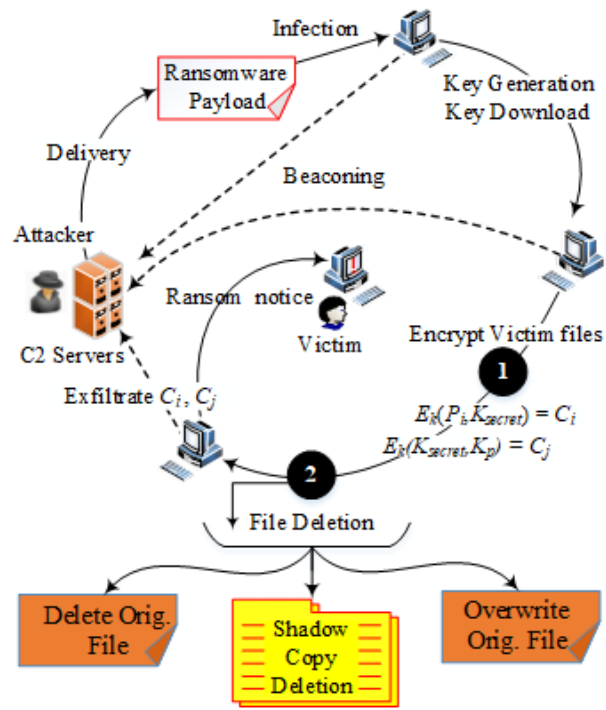

Figure 3. Threat model and attack structures

The model comprises the attacker with command and control (C2) server resources who seeks to victim a host. The C2 might house the ransomware or associated encryption keys depending on the attack model. The attacker uses any of the discussed cryptosystems for encryption and chooses an effective infection vector [13]. Depending on the ransomware variant, the attacker might embed the encryption key in the ransomware payload or the ransomware would have to beacon to the C2 upon infection to download the necessary encryption keys. To effectuate an effective ransomware attack, the ransomware carries out two major tasks; (1) encrypt the target files and (2) delete the original files after encryption. Encryption of the target files is denoted by $E_k(P_i, K_{secret}) = C_i$. Some attack structures generate a symmetric key using the victim's operating system *CryptoAPI* [14]. After this key completes encrypting the target files, it is further encrypted by the embedded public key which is denoted by $E_k(K_{secret}, K_p) = C_j$. The resultant ciphertext $C_j$ is exfiltrated to the C2 server. In the case of single key attack model, the encryption process is denoted as:

$$\{m_i(target\_payload)\}_{K_{pub}} \rightarrow C_i \quad (1)$$

$$\{m_i(target\_payload)\}_{K_{secret}} \rightarrow C_i \quad (2)$$

Equation (1) is an implementation of an asymmetric cryptosystem whilst Equation (2) a symmetric cryptosystem. In the case of a hybrid key attack model, the encryption attack process is denoted as:

$$\{m_i(data_1)\}_{K_{sym}} \rightarrow C_i \quad (3)$$

$$\{K_{secret}(data_2)\}_{K_{pub}} \rightarrow C_j \quad (4)$$

Equation (3) denotes the first stage of the encryption process where $m_i$ is the plaintext message (targeted files) $data_1$ and $C_i$ is the resultant ciphertext. Equation (4) is the second stage of the encryption process where $data_2$ is the symmetric key $K_{secret}$ used in Equation (3) and $K_{pub}$ is the public key while $C_j$ is the resultant ciphertext.

After completion of the encryption, the ransomware proceeds to delete the remnant files after encryption and the volume shadow copies. The volume shadow copies are usually deleted via vssadmin.exe while the remnant files are either deleted primitively by erasing directories structures and meta-data information of the files. The other way of deleting the files is by overwriting it with random data which corrupts the file and make it unreadable. If the files are deleted via meta-data information and directories structures, they are easily recoverable via third party software and utilities. On the other hand, overwriting the target file with random data makes recovery very difficult.

Considering the above attack structures (file encryption and file deletion), we propose a ransomware classification framework that is based on evaluation of the underlying attack structures. This is helpful because poorly implemented attack structures make recovery of data possible regardless of the resilience of the used crypto-algorithms.

### III. PROPOSED CLASSIFICATION FRAMEWORK

We propose a ransomware classification framework that is based on the attack structures depicted in the threat model in figure 3 and on the characteristics from the Cryptosystem-Based and Attack Structure-Based categories in the taxonomy in figure 2. The classification framework is shown in Table 2.

We use this categorization framework to formulate a classification algorithm that classifies a ransomware given its attack structures. The algorithm is shown in figure 4 below.

Algorithm: Ransomware Classification

```
Input: Encryption & deletion attack structures
Output: Ransomware category
1.  if SKc2emb=SKPemb=SKlocalgen=
        HKc2emb=HKPemb=HKlocalgen=no then
2.      malware ← CAT1
3.  else
4.      if delShdCpy=ovrFile=no then
5.          malware ← CAT2
6.      else
7.          if SKc2emb=SKPemb=SKlocalgen=no then
8.              malware ← CAT5
9.          else
10.             if SKc2embsym = SKPembsym =
                    SKlocalgensym = yes then
11.                 malware ← CAT3
12.             else
13.                 malware ← CAT4
14.             end if
15.         end if
16.     end if
17. end if=0
```

Figure 4. Ransomware virulence classification algorithm

**Table 1**. Abbreviations of algorithm parameters

| Feature | Term | Code |
|---|---|---|
| Hybrid cryptosystem | C2 download | HKc2emb |
| | Payload embedded | HKPemb |
| | Local key Generation | HKlocalgen |
| Single Key Cryptosystem | C2 download | SKc2emb |
| | Payload embedded | SKPemb |
| | Local key Generation | SKlocalgen |
| Delete Volume Shadow Copies | | delShdCpy |
| Overwrite & Delete Original File | | ovrFile |

Our framework expresses the severity of a ransomware in terms of file encryption and file deletion. As such, it shows how challenging and time consuming it will be to mitigate a given ransomware attack using the classical methods of static and dynamic analysis [15]. The virulence depicted in the framework is flexible, i.e. a ransomware can move up or down the category list depending on newly discovered properties.

Since the framework categorizes the ransomware in ascending order, it is clear that ransomware CAT1 is easier to mitigate than CAT5. As such, CAT5 is the most virulent whilst CAT1 is the least virulent where recovery of data does not require any decryption keys. Since the first sub-category of CAT1 does not implement any of the earlier discussed attack structures, the severity is negligent and it's thus categorized as Scareware as depicted in the taxonomy in figure 2. Examples of such malware include AnonPop [16].

Table 2. Ransomware Classification Framework

| CATEGORY (Severity) | ENCRYPTION ATTACK MODEL | | | | | | DELETION ATTACK MODEL | |
| --- | --- | --- | --- | --- | --- | --- | --- | --- |
| | Hybrid cryptosystem | | | Single Key Cryptosystem | | | Delete Volume Shadow Copies | Overwrite & Delete Original File |
| | C2 download | Payload embedded | Local key Generation | C2 download | Payload embedded | Local key Generation | | |
| CAT1 | ✗ | ✗ | ✗ | ✗ | ✗ | ✗ | NO | NO |
| | ✗ | ✗ | ✗ | ✗ | ✗ | ✗ | YES | NO |
| | ✗ | ✗ | ✗ | ✗ | ✗ | ✗ | NO | YES |
| | ✗ | ✗ | ✗ | ✗ | ✗ | ✗ | YES | YES |
| CAT2 | ✓ \|\| ✓ \|\| ✓ | | | ✗ | | | NO | NO |
| | ✗ | | | ✓ \|\| ✓ \|\| ✓ | | | NO | NO |
| CAT3 | ✗ | ✗ | ✗ | ✓ ($K_{enc} = K_{sym}$) | | | YES | YES |
| CAT4 | ✗ | ✗ | ✗ | ✓ \|\| ✓ \|\| ✓ | | | YES | YES |
| CAT5 | ✓ \|\| ✓ \|\| ✓ | | | ✗ | ✗ | ✗ | YES | YES |

Other sub-categories in CAT1 implement some of file deletion but not encryption. Thus, recovery of data is possible via third-party software such as Recuva or Photorec [17]. In all these cases, data is recoverable without the need to honor the ransom demand. As such, mitigation measures should never focus on key retrieval as there's no key in the attack structure. CAT2 ransomware employs only the file encryption attack structures. The key can be download from the C2 or it can come embedded in the payload. This is an example of a poorly implemented ransomware as was the case with Bad Rabbit [18] despite using robust cryptosystems. Since there's no deletion of volume shadow copies, data can be recovered via system restore utilities or third-party software. CAT3 represents earlier and uncommon types of ransomware that are based on single key attack structures. In this category, the ransomware comes with an embedded symmetric key in the payload. The key can be simply retrieved using reverse engineering. In the event that the key is deleted from the payload, data deletion recovery techniques discussed in the preceding categories can be used to recover the key. However, if the embedded key is a public key from an asymmetric cryptosystem, it is of no value to extract the public key since it cannot decrypt the data. This is representative of CAT4. Another instance of CAT4 is where the key is downloaded from the C2 server. The key can be symmetric or asymmetric as was with the case of CryptoWall [19]. In the case of the latter, it is very difficult to mitigate the attack since there are no residual encryption essentials on the victim. CAT5 represents the current generation of ransomware. The attack structures implement all the deletion techniques and use hybrid cryptosystems. A typical example is Wannacry which deletes not only the volume shadow copies but the remnant files as well. Further, it comes with an embedded master RSA public key and uses the operating system's CryptoAPI to generate an RSA sub-key pair and AES keys. Each of the unique AES keys is used to encrypt a unique target file. The embedded RSA master public key is used to encrypt the private key from the generated RSA sub-key pair. The public key of the generated RSA sub-key pair is used to encrypt the unique AES keys. As such, to decrypt the data, the victim needs the AES which is encrypted by the private key of the generated RSA sub-key pair. This can only be decrypted by the corresponding generated RSA private sub-key pair, but then, it has been encrypted by the embedded RSA master public key. What the RSA master key has encrypted can only be decrypted by the corresponding RSA master private key which is in the domain of the attacker. The attacker thus demands a ransom to release the decryption key.

IV. METHODOLOGY AND APPROACH

To evaluate the feasibility and effectiveness of our framework, we analyzed 20 ransomware samples and applied the algorithm of the framework for categorization purposes as shown in figure 5.

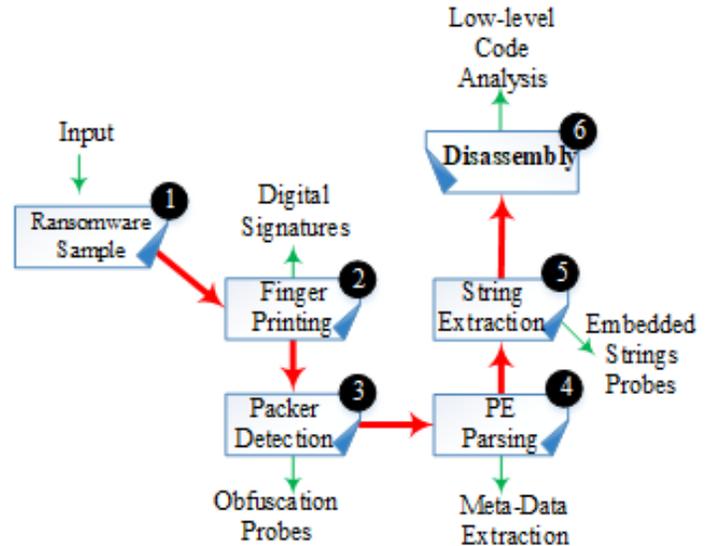

Figure 5. Static analysis workflow

We analyze and extract encryption and deletion features from the ransomware using two approaches: static code analysis and behavioral analysis. The workflow for static analysis is shown in figure 5. In step 1, we choose the ransomware binaries from reputed malware sources not limited to ReverseIT, VirusTotal and Malware Byte's Malwr. We verify the malware's fingerprints by computing corresponding cryptographic hashes in step 2. We check whether the malware is obfuscated via packing in step 3 and then parse it for meta-data extraction in step 4. We probe any embedded strings in step 5 and finally analyze the low-level code in step 6. We mainly use IDA Pro and Ollydebug complemented with PEView.

Furthermore, we complement static analysis with behavioral analysis. We add this extra step because there are some ransomware attack-structure features that only appear when the malware is actively executing. We use Cuckoo sandbox as our execution environment. The Cuckoo server runs on Kali Linux whist the virtual hosts through which we actively monitor the malware's activities run Windows on VirtualBox. Though we run the test-bed in *host-only-adapter*, we simulate the Internet by sink-holing it with the *FakeDNS* utility.

By thoroughly running through the processes of static and behavioral analysis, we extract attack structure features which we use as input for the algorithm in figure 4. As such, we classify the virulence of the ransomware according to the formulated categories. The results of the classification are presented in the next section.

## V. CLASSIFICATION RESULTS AND ANALYSIS

The results from the analyzed samples and their corresponding categories are shown in Table 3 below. The *Name* column denotes the name of a the particular strain of the malware. We stick to the initial name associated with ransomware and not its subsequent versions. The *Year* depicts the period during which the given sample version existed. It is worth noting that most of the ransomware activities span a couple of years and during this period, newer versions appear with enhanced capabilities. For example, the earliest version of DMA-Locker was seen in 2015 and it used a single symmetric key (AES-256 in ECB mode) to encrypt all the targeted file. And since it deleted remnant files and volume shadow copies, it falls into CAT4. However, in 2016, the enhanced version appeared which used a separate AES key for encryption of each file [24]. Furthermore, the used AES keys were encrypted by an embedded RSA public key. Based on our classification framework, this essentially moves DMA-Locker from CAT4 to CAT5. As such, the *Year* column in table 3 depicts the particular year in which the associated malware variant was seen. The *Paid Ransom* column attaches the monetary value associated with each ransomware campaign. The null entries depict unavailability of verified data. Furthermore, the monetary value associated with each campaign might be more because some ransomware variants are known to use multiple Bitcoin addresses. Some other variants have been known to accept payments in other forms of digital money other than Bitcoins [21] implying that the cumulative value is more than those we traced from BlockChain Info [22]. The *Platform* column denotes the targeted operating system of which 85% represents Windows. The *Category* shows the class in which the corresponding ransomware falls. As can be seen from Table 3, most of the ransomware variants fall into CAT4 and CAT5 accounting for 35% each. CAT1 is not common and is usually the work of script-kiddies whilst CAT2 and CAT3 encompasses early unmatured variants of ransomware such as AIDS [23]. Furthermore, ransomware in CAT4 and CAT5 account for the highest values in paid ransoms. This can be attributed to the difficulty in mitigating such ransomware categories owing to the complex encryption and deletion attack structures.

**Table 3**. Classification of notable ransomware incidents

| Name | Year | Paid Ransoms | Platform | Category |
|---|---|---|---|---|
| AnonPop | 2016 | - | Windows | *CAT1* |
| Cerber | 2016 | > $500,000 | Windows | *CAT5* |
| Bad Rabbit | 2017 | - | Windows | *CAT2* |
| CryptoDefense | 2014 | > $65,000 | Windows | *CAT4* |
| CryptoLocker | 2014 | > $ 3 million | Windows | *CAT4* |
| CryptoWall | 2015 | $18 million | Windows | *CAT4* |
| DMA-Locker | 2015 | > $180,000 | Windows | *CAT4* |
| Jigsaw | 2016 | > $2,000 | Windows | *CAT3* |
| Erebus | 2017 | > $1.04 million | Linux | *CAT5* |
| NotPetya | 2017 | > $10,000 | Windows | *CAT3* |
| KeRanger | 2016 | > $5,000 | Mac OS | *CAT4* |
| Linux.Encoder | 2015 | - | Linux | *CAT3* |
| Locky | 2016 | > $ 1.3 million | Windows | *CAT4* |
| AIDS | 1989 | - | Windows | *CAT3* |
| Petya | 2016 | > $30,000 | Windows | *CAT5* |
| SamSam | 2018 | > $850,000 | Windows | *CAT5* |
| TeslaCrypt | 2015 | > $80,000 | Windows | *CAT4* |
| VenusLocker | 2016 | > $6,500 | Windows | *CAT5* |
| WannaCry | 2017 | > $140,000 | Windows | *CAT5* |
| ZCryptor | 2016 | - | Windows | *CAT5* |

The evolution of ransomware file-deletion and file-encryption characteristics has seen an increment in the emergence of new resilient ransomware mostly falling in CAT5 as depicted in ransomware-attack statistics in figure 6 [20]. The surge in ransomware attacks represent a 229% increment most of which are CAT4 and CAT5 ransomware.

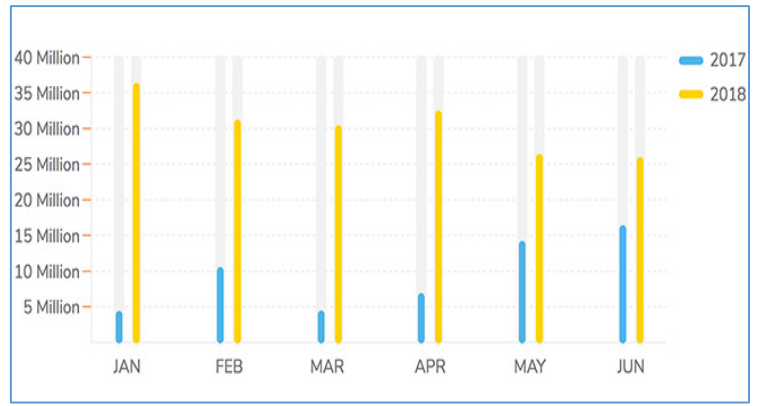

Figure 6. Global ransomware attacks volume for 2017/2018

The changes in ransomware attack structures are echoed in figure 7 below based on our data-set. Poorly designed ransomwares are neglected over the years as was the case with CAT2 ransomware which appear only in 2017 and not prior or after. This is a common characteristic erroneously or poorly implemented ransomware variants.

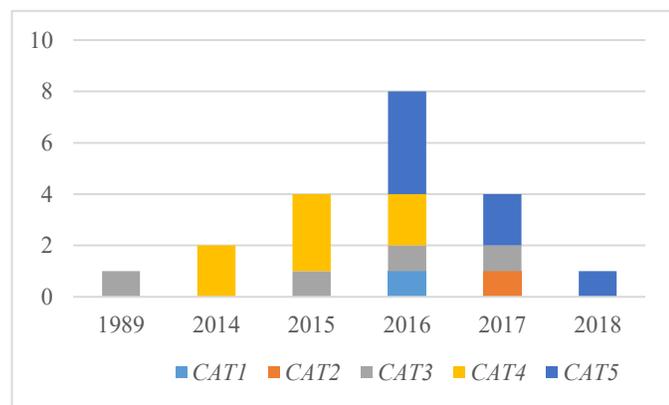

Figure 7. Distribution of ransomware categories over the years

The years 2014 – 2016 see a steady appearance of CAT4 ransomware which is followed by a steady appearance of CAT5 ransomware from 2016 – 2018. Other ransomwares are resilient and span several years. This is the case with CAT3 ransomware except for the AIDS ransomware of 1989. By volume, it is evident from figure 7 that CAT4 and CAT5 are the most common followed by CAT3. CAT3 and CAT4 ransomware can be mitigated effectively via reverse engineering (static analysis) provided the key used is symmetric. CAT5 can be mitigated if the encryption attack structure uses hybrid encryption essentials from the victim.

## VI. CONCLUSIONS

In this paper, we have presented a ransomware classification framework based on file-deletion and file-encryption attack structures. Based on these two attack structures, we have presented a thorough taxonomy of ransomware attacks which formed the basis of our classification framework. We have defined the threat model and attack structures to characterize a detailed overview of the ransomware attack process not only in terms of encryption but data deletion as well. The threat and attack models are representative of a typical ransomware attack process from which we derived the ransomware categorization framework based on a proposed classification algorithm. The framework classifies the virulence of a ransomware attack to entail the overall effectiveness of potential ways of recovering the attacked data without paying the ransom demand as well as the technical prowess of the underlying attack structures. The algorithm classifies the virulence of a ransomware in increasing severity from CAT1 through to CAT5. We carried out static and dynamic malware analysis to extract the features for use in the framework. Results show that the most recent ransomware strains are CAT4 and CAT5 which are better mitigated by exploiting encryption essentials. CAT1 and CAT2 ransomware are not common in the wild whilst CAT3 and CAT4 ransomware can be mitigated effectively via reverse engineering (static analysis) provided the key used is symmetric. CAT5 can be mitigated if the encryption attack structure uses hybrid encryption essentials from the victim.